# Vibration-suppressed toolpath generation for kinematic and energy performance optimization in large dimension additive manufacturing

Kang WANG [a*] Jinghua XU [a] Shuyou ZHANG [a] Jianrong TAN [a]

[a] School of Mechanical Engineering, Zhejiang University, Hangzhou, 310027, China

* Corresponding author

**Abstract:** Product quality and sustainability in large dimension additive manufacturing (LDAM) highly rely on the kinematic and energy performance of LDAM systems. However, the mechanical vibration occurring in the fabrication process hinders the further development of AM technology in industrial manufacturing. To this end, a kinematic and energy performance optimization method for LDAM via vibration-suppressed toolpath generation (VTG) is put forward. Kinematic modeling of LDAM system and related energy consumption modeling are first constructed to reveal that the toolpath of servo system assumes the main responsibility for kinematic and energy performance in LDAM. The vibration-suppressed toolpath generation (VTG) method is thus proposed to customize servo trajectory for kinematic and energy performance optimization in LDAM. Extensive numerical and physical experiments are conducted on the W16 cylinder, including the measurement and analysis of the mechanical vibration amplitude, and assessment of the surface roughness before and after using the proposed VGT. These experimental results demonstrate that our VGT is able to simultaneously achieve kinematic and energy performance optimization in LDAM, even though the fabricated part owns aberrant and complex morphology.

1. **Introduction**

Additive manufacturing (AM) is an intelligent fabrication technology, which uses CNC (Computerized Numerical Control) system to produce end-use parts by directly using the computer aided design (CAD) model without any auxiliary tools such as dies [1,2]. Given a three-dimensional (3D) CAD model, AM slices the model to multiple two-dimensional (2D) layers and fabricates these layers layer-upon-layer, decreasing the difficulty of fabricating parts with complex structure [3–5]. This layer-wise fabrication manner enables AM fabrication capability not limited to the product design process, and hence AM technology has drawn the attention from academics and industries since its advent [6,7].

Major end-user market applications and industries that benefit from 3D printing include aerospace, automotive, medical and dental, consumer goods, hobbies and personal use [8–11], to mention a few. However, current 3D printer market is dominated by small dimension 3D printers (commonly known as desktop 3D printers) with an average approximate workspace of 200 × 200 × 200 mm [12]. The limited printing volume of small dimension 3D printers hinders the further development of AM technology in industrial manufacturing [13,14]. As a consequence, there exists the increasingly demand to exploit large dimension additive manufacturing (LDAM), enabling to upscale the physical dimensions of AM products.

To optimize the fabrication efficiency and quality, a variety of researches have been undertaken for LDAM from various perspectives, including process planning [15–17], support structure design [18,19], and mechanical improvement [20,21]. For example, Diourté et al. [16] proposed a wire arc additive manufacturing strategy to minimize the start/stop phases of the arc to one unique cycle, by generating a continuous trajectory in spiral form for closed-loop thin parts. Allum et al. [17] proposed a non-planar geometries, called ZigZagZ, which deposits filaments simultaneously in the X, Y, Z direction. Jiang et al. [22] proposed a support generation method to reduce support material consumption by considering the effect of the longest printable bridge length to support usage. Delfs et al. [23] developed a prediction method of the surface quality in dependence of the building orientation of a part, and the prediction results are thus be used to optimize the orientation to get a desired surface quality.

Despite these optimizations of the fabrication efficiency and quality having remarkable success, the kinematic performances of LDAM still suffer from the mechanical vibration during the manufacturing process [24,25]. To analyze the mechanical vibration in AM system, Wu et al. proposed a new method for in situ monitoring of FDM machine conditions where the acoustic emission technique is applied [26]. Tlegenov et al. proposed a nozzle condition monitoring technique in fused filament fabrication 3D printing using a vibration sensor [27]. Zhang et al. employed an attitude sensor to monitor signals of multiple sensors mounted on the moving platform of the printer by using error fusion of multiple sparse auto-encoders [28]. However, most of the existing methods are unable to fundamentally solve the vibration problem in LDAM, since the vibration is an inherent attribute of mechanical process due to the intrinsic kinematic system, but only could be suppressed.

To circumvent this difficulty, recent studies have shown that toolpath planning becomes a promising method to provide optimal trajectory for kinematic system, suppressing the mechanical vibration in fundamental respects for LDAM. Dwivedi and Kovacevic [29] proposed an automated continuous toolpath planning approach, which connects individual zigzag paths in each decomposed infill sub-region. Lin et al. [30] introduced dual offset and clear-up toolpath procedures to alter the path, aiming to alleviate the problems from sharp corners. Griffiths [31] proposed a Hilbert's curve based toolpath strategy for machining curved surfaces. Kuipers et al. [32] proposed a adaptive width control strategy by using various beading schemes to reduce the bead width range. Etienne et al. [33] developed a curved slicing method to better align the object surface by generating a toolpath to adapt the variable slicing layer height. The mentioned-above researches contribute to the favorable development of AM technologies. However, there are limited researches to exploit the mechanical vibration characteristics of LDAM system, though they determine the final kinematic performance and surface quality when the physical dimension of AM machine becomes larger. In addition, most of toolpath planning methods are validated on the art crafts with relatively ordinary geometry, which may be not applied in industrial product with complex structures.

Aiming to dynamically improve kinematic and energy efficiency performance, this paper proposes a vibration-suppressed toolpath generation (VTG) method for kinematic and energy performance optimization in large dimension additive manufacturing. The proposed VTG is verified by a series of coherent numerical and physical experiments on a W16 cylinder which owns aberrant and complex morphology features.

The main contributions of this paper are summarized as follows:

- A Kinematic modeling of LDAM system is constructed by using the equivalent lumped mass model, to reveal the relationship between the input force and output displacements in LDAM system.
- The energy consumption modeling related to LDAM kinematics is put forward, signifying that the energy of servo system accounts for main part of the total energy consumption in LDAM.
- In light of the above-established kinematic model and the corresponding energy consumption model, the vibration-suppressed toolpath generation (VTG) method is proposed to customize servo trajectory for kinematic and energy performance optimization in LDAM.
- Extensive numerical and physical experiments are conducted on the W16 cylinder, demonstrating that our VGT is able to simultaneously achieve kinematic and energy performance improvement in LDAM.

## 2. Preliminary

### 2.1 Kinematic modeling of LDAM system

Fig. 1 shows the kinematic system of the LDAM machine in XYZ axis format, where the 3D printer has three vibration sources from extruder, servo system, and auxiliary fan, respectively. Obviously, the working state of ball screw determines the kinematic performance of LDAM system. Any vibration from each component can be passed to working table by the ball screws, and severe mechanical vibration of LDAM system induces dislocation and printing fault, directly leading to the unsatisfied product quality. The frequent vibration of LDAM system also greatly reduces the service life of the ball screw. Consequently, it is of great significance to establish a dynamic kinematic model, revealing the relationship between the input force and output displacements in LDAM system.

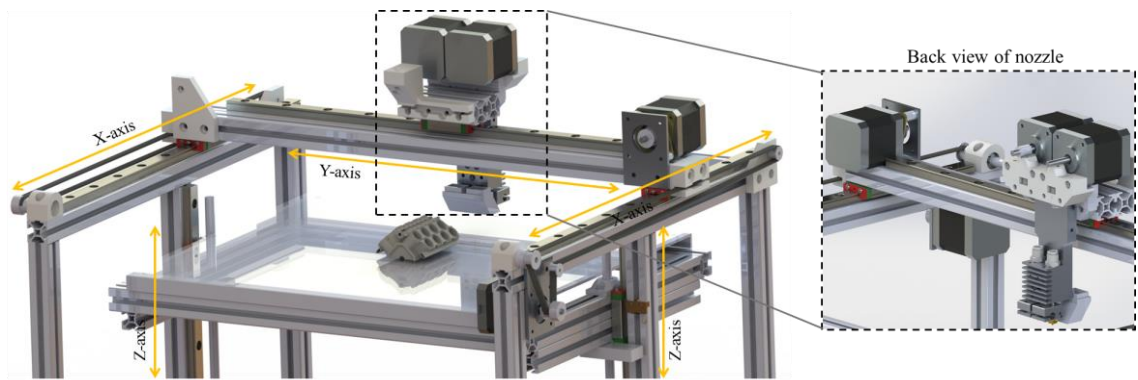

**Fig. 1** Kinematic system of the LDAM machine in XYZ axis format.

As shown in Fig. 2, the mechanical system of LDAM is composed by electric servo-motor, ball screw, screw nut pair, flexible coupling, fixed bearing at both ends, working table, and sliding rail. To analyze the mechanical system of LDAM, the kinematic modeling of LDAM system is firstly constructed by using the notion of equivalent lumped mass.

Since the ball screw is connected with electric servo-motor by the coupling, the equivalent mass of the ball screw takes them both into account. There are two similar ball screws including ball screw 1 and 2. For brevity, the ball screw 1 is considered as the example, and its equivalent mass $m_1$ is represented as follow:

$$m_1 = m_{sc1} + m_{r1} + m_{c1} \tag{1}$$

$$m_{sc1} = \pi \rho l_s r_s^2 \tag{2}$$

where $m_{sc1}$ is ball screw mass (kg), $m_{r1}$ is internal rotor mass (kg), $m_{c1}$ is coupling mass (kg), $\rho$ is material density (kg·m⁻³), $l_s$ is ball screw length (m), and $r_s$ is ball screw radius (m).

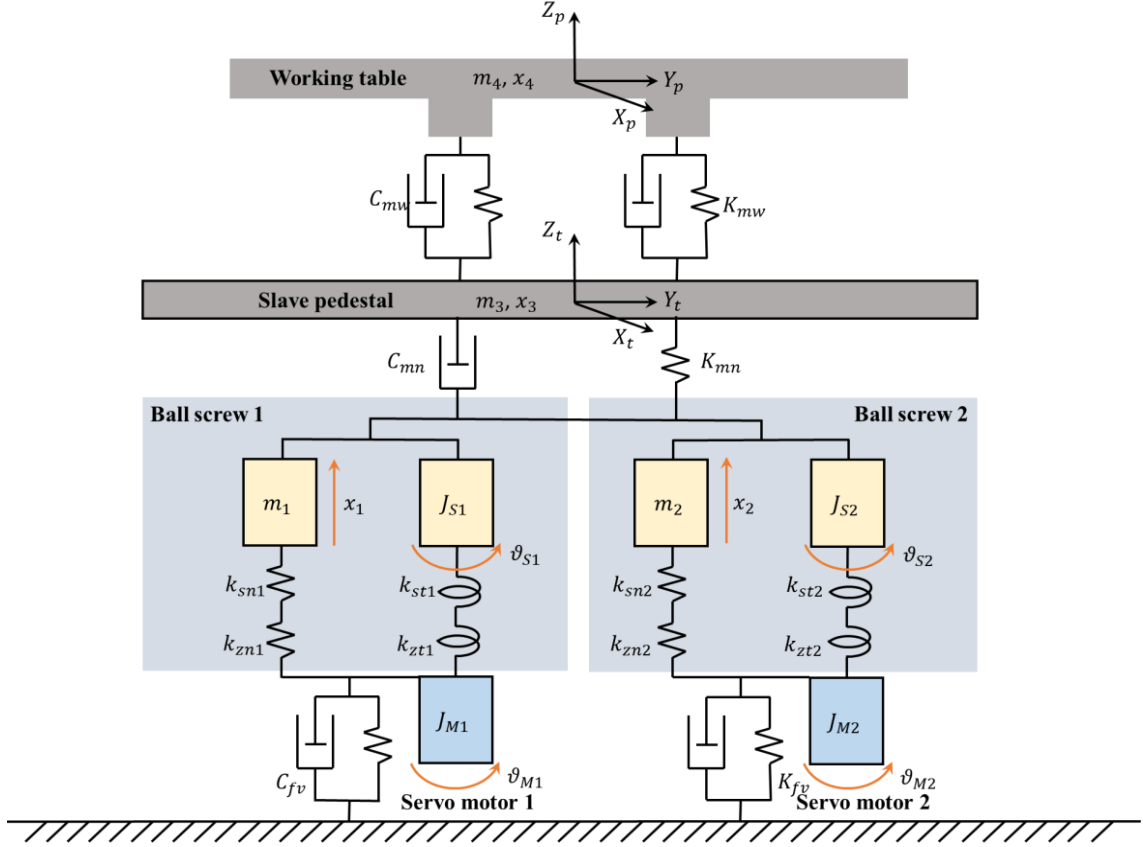

**Fig. 2** Schematic of the equivalent lumped mass modeling for the LDAM machine.

The equivalent inertia moment of ball screw 1 is represented as follow:

$$J_{S1} = J_{sc1} + J_{c1} + i_{s1}^2 m_3 \tag{3}$$

$$J_{sc1} = \frac{\pi \rho l_s r_s^4}{4} \tag{4}$$

where $J_{S1}$ (kg·m²) is equivalent inertia moment of ball screw 1, $J_{sc1}$ (kg·m²) is inertia moment of the ball screw, $J_{c1}$ (kg·m²) is inertia moment of the coupling, $m_3$ (kg) is slave pedestal mass, and $i_s$ is the transmission ratio of the ball screw.

The axial stiffness and torsional stiffness of ball screw 1 are related to the position of nut on the slave pedestal, and their definitions are described as below:

$$k_{sn1} = \pi r_s^2 \frac{E}{l_n} \tag{5}$$

$$k_{st1} = \frac{G\pi r_s^2}{4l_n} \tag{6}$$

where $k_{sn1}$ (N·m⁻¹) is axial stiffness of ball screw 1, $E$ (MPa) is the elastic modulus of ball screw material, $l_n$ (m) is the length between of nut and coupling, $k_{st1}$ (N·m⁻¹) is torsional stiffness, and $G$ (MPa) is the shear modulus of ball screw material.

There are multiple ball screw nuts in the ball screw pair to transmit the torque to linear movement. Taking the nut between the ball screw and slave pedestal as the example, its contact stiffness $k_{mn}$ (N·m⁻¹) is obtained as below:

$$k_{mn} = 0.8 k_c \left(\frac{F_z}{0.1 C_a}\right)^{\frac{1}{3}} \tag{7}$$

where $k_c$ (N·m⁻¹) is the reference stiffness without preloading force, $F_z$ (N) is the level of preloading force, and $C_a$ (N) is the dynamic load rating.

In addition, the mechanical system of LDAM also requires the fixed bearing to support the ball screw. To combine the axial stiffness of the bearing and ball screw, the axial composite stiffness $k_{n1}$ (N·m⁻¹) of bearing and ball screw 1 is obtained in term of the elasticity theory principle as below:

$$k_{n1} = \left(\frac{1}{k_{zn1}} + \frac{1}{k_{sn1}}\right)^{-1} \tag{8}$$

where $k_{zn1}$ is axial stiffness of the bearing for ball screw 1, and $k_{sn1}$ (N·m⁻¹) is axial stiffness of ball screw 1.

Similarly, torsional composite stiffness $k_{t1}$ (N·m⁻¹) of ball screw 1 and coupling can be obtained as below:

$$k_{t1} = \left(\frac{1}{k_{zt1}} + \frac{1}{k_{st1}}\right)^{-1} \tag{9}$$

In Fig. 2, $\theta_M, \theta_S$ (rad) denote equivalent angular displacements of two torsion spring, respectively. $x_1, x_2, x_3, x_4$ (m) denote equivalent axial displacements of four tension spring, respectively. These displacements constitute the LDAM system freedom degrees $q = [x_1, x_2, x_3, x_4, \theta_M \ \theta_S]^T$ as the generalized coordinates. The Lagrange equation of the mechanical system of LDAM is then established as below:

$$\frac{d}{dt}\left(\frac{\partial E_k}{\partial \dot{q}_i}\right) - \frac{\partial E_k}{\partial q_i} + \frac{\partial D}{\partial \dot{q}_i} + \frac{\partial E_p}{\partial q_i} = F_i \quad (i = 1,2,\cdots,n) \quad (10)$$

where $E_k, E_p$ (J) are respectively total kinetic energy and total potential energy of the system, $i$ is the number of generalized coordinates in the LDAM system, $q_i$ is a generalized coordinate, $D$ is the power dissipation in the system caused by the viscous damper, and $F_i$ is generalized force.

Formally, total kinetic energy $E_k$, and total potential energy $E_p$ of the LDAM system are respectively obtained by the Kinetic Energy Law as follows:

$$E_k = \frac{1}{2}\left(m_1\dot{x}_1^2 + m_2\dot{x}_2^2 + m_3\dot{x}_3^2 + m_4\dot{x}_4^2 + J_M\dot{\theta}_M^2 + J_S\dot{\theta}_S^2\right) \quad (11)$$

$$E_p = \frac{1}{2}k_n x_1^2 + \frac{1}{2}k_t(\theta_M + \theta_S)^2 + \frac{1}{2}k_{mn}(x_3 - x_s - i_s\theta_S)^2 + k_{mw}(x_4 - x_3)^2 \quad (12)$$

$$D = C_{mn}\dot{x}_3^2 + 2C_{mw}\dot{x}_4^2 \quad (13)$$

where $J_M, J_S$ are the equivalent inertia moment of the motor and the ball screw respectively, $m_1, m_2, m_3, m_4$ are the equivalent mass of ball screw 1, ball screw 2, slave pedestal, and working table respectively, $x_s$ is the average displacement of ball screws defined by $x_1$ and $x_2$, $k_{mn}$ is contact stiffness of the ball screw nut, $k_{mw}$ is axial stiffness of flexure hinge, and $C_{mn}, C_{mw}$ are two different damper coefficients of ball screw nut and flexure hinge respectively.

To quantize the vibration displacement, the kinematic model of mechanical vibration for LDAM system can be obtained according to the Eq. 11 as below:

$$\boldsymbol{M\ddot{q} + C\dot{q} + Kq = F} \quad (14)$$

where $\boldsymbol{F}$ is the external load (N), and $\boldsymbol{M}$, $\boldsymbol{C}$, and $\boldsymbol{K}$ are the mass, damping, and stiffness matrices of LDAM system, respectively.

Let $C(x,y,z)$ $C(x_p,y_p,z_p)$ denote the theoretical and measured displacement of working table displacement $x_4$ in XYZ-axis. To evaluate the mechanical vibration amplitude, the trajectory deviation $dev$ is then represented as below:

$$dev = \|C(x,y,z) - C(x_p,y_p,z_p)\| \quad (15)$$

where $\|.\|$ is the operator to calculate 2-norm value.

## 2.2 Energy consumption modeling related to LDAM kinematics

The unexpected vibration naturally causes the extra energy consumption according the Kinetic Energy Law. To analyze the energy performance of LDAM system, the total energy consumption $E$ can be divided into three parts as below:

$$E = E_{servo} + E_{fuse} + E_{aux} \qquad (16)$$

where $E_{servo}$ is the energy of servo system (J), $E_{fuse}$ is the energy to fuse thermoplastic material (J), and $E_{aux}$ is the energy for auxiliary tools (J).

To quantitatively analyze the impact of energy consumption in LDAM on the environment, the mass of the carbon emission $m_{CO_2}$ (kg $CO_2$) is computed as below:

$$m_{CO_2} = \overline{f_{CO_2,e}} * E + \overline{f_{CO_2,m}} * m_{mater} \qquad (17)$$

where $\overline{f_{CO_2,e}}$ is average carbon emission of unit energy consumption (kg $CO_2 \cdot kw^{-1} \cdot h^{-1}$), and $\overline{f_{CO_2,m}}$ is average carbon emission of unit material (kg $CO_2 \cdot kg^{-1}$).

According to the above-established kinematic modeling and related energy consumption modeling, it obviously signifies that the energy of servo system $E_{servo}$ accounts for main part of the total energy consumption. It thus inspires us to design a vibration-suppressed toolpath generation to synchronously optimize kinematic and energy performance in LDAM.

## 3. Methodology for optimizing kinematic and energy performance

### 3.1 Gaussian curvature and NURBS

To evaluate the morphology of CAD model, Gaussian curvature $K$ could be an effective metric to denote bending degree of an arbitrary slicing curve. The first and second derivatives at any point on the surface $\boldsymbol{S}(u,v)$ can be calculated according to the differential geometry theory, which are denoted as follows:

$$S_u = \frac{\partial \boldsymbol{S}}{\partial u}, \quad S_v = \frac{\partial \boldsymbol{S}}{\partial v}, \quad S_{uu} = \frac{\partial^2 \boldsymbol{S}}{\partial u^2}, \quad S_{vv} = \frac{\partial^2 \boldsymbol{S}}{\partial v^2}, \quad S_{uv} = \frac{\partial^2 \boldsymbol{S}}{\partial u \partial v} \qquad (18)$$

Given a point on the surface $S(u, v)$, it has countless curves and corresponding normal curvatures. Among these curvatures, maximum curvature $K_{max}$ represents the maximum normal curvature. Minimum $K_{min}$ represents the normal curvature perpendicular to $K_{max}$. Also, $K_{max}$ and $K_{min}$ are termed as two principal curvatures. The mean curvature $H$ and Gaussian curvature $K$ is then obtained by the principal curvatures as follows:

$$H = \frac{1}{2}(k_{min} + k_{max}) = \frac{LG - 2MF + NE}{2(EG - F^2)} \tag{19}$$

$$K = k_{min} k_{max} = \frac{LN - M^2}{EG - F^2} \tag{20}$$

where $E = S_u \cdot S_u$、$F = S_u \cdot S_v$、$G = S_v \cdot S_v$ are the first basic quantity of surface, $\boldsymbol{n}$ is the unit normal vector on surface, $L = \boldsymbol{n} \cdot S_{uu}$、$M = \boldsymbol{n} \cdot S_{uv}$、$N = \boldsymbol{n} \cdot S_{vv}$ are the second basic quantity of surface.

To represent slicing curves in LDAM, Non-uniform rational basis spline (NURBS) is adopted to fit and optimize the generated toolpath, which is a commonly mathematical model in computer graphics. The NURBS curve in $p$-order can be expressed as:

$$C(u) = C(x, y, z) = \frac{\sum_{i=0}^{n} N_{i,p}(u) P_i w_i}{\sum_{i=0}^{n} N_{i,p}(u) w_i} \tag{21}$$

$$N_{i,p}(u) = \begin{cases} N_{i,0}(u) = \begin{cases} 0, u_i \leq u \leq u_{i+1}; \\ 1, \quad else\ ; \end{cases} \\ N_{i,p}(u) = \frac{u - u_i}{u_{i+p} - u_i} N_{i,p-1}(u) + \frac{u_{i+p+1} - u}{u_{i+p+1} - u_{i+1}} N_{i+1,p-1}(u) \end{cases} \tag{22}$$

where $C(u)$ is NURBS curve, $u \in \{u_0, u_1, \cdots, u_{n+p+1}\}$ is knot vector, $w_i \in \{w_0, w_1, \cdots, w_n\}$ is weights vector, $p$ is the order of $C(u)$, $P_i \in \{p_0, p_1, \cdots, p_n\}$ is control points vector, $N_{i,p}(u)$ is $p$-order B-spline basis function obtained by the *Cox-de Boor* recursion.

In addition, normalized height $h_n$ is defined to describe the precise slicing layer position as below:

$$h_n = \frac{Z_i}{Z} \times 100\% \tag{23}$$

where $Z_i$ is layer height of the $i$-th slicing layer, $Z$ is total height of the AM product.

## 3.2 Vibration-suppressed toolpath generation (VTG) considering relations between kinematic and energy performance

In light of the above-established kinematic model and the corresponding energy consumption model, the vibration-suppressed toolpath generation (VTG) method is proposed to customize servo trajectory for kinematic and energy performance optimization in LDAM. The proposed VTG takes trajectory length and smoothness into account, enabling to adapt to the complex morphology of manufactured product.

Let $n_{total}, n_{turning}, n_{infill}$ denote the number of total trajectory points, turning trajectory points, and infill trajectory points, respectively. Let $L_{total}, L_{infill}$ respectively denote length of total trajectory points, and infill points that are the majority type of total trajectory points, which is represented as below:

$$L_{total} = \sum_{i=1}^{n_{total}-1} \|C_{i+1}(x,y,z) - C_i(x,y,z)\| = \sum_{j=1}^{n_L} L_{total}^j \qquad (24)$$

$$L_{infill} = \sum_{i=1}^{n_{infill}-1} \|C'_{i+1}(x,y,z) - C'_i(x,y,z)\| \qquad (25)$$

where $C_i(x,y,z)$ is the $i$-th trajectory point, $C'_i(x,y,z)$ is the $i$-th infill trajectory point, $L_{total}^j$ is trajectory length of the $j$-th slicing layer.

To evaluate the trajectory smoothness, the turning points are defined by turning angle $\vartheta_i$ as below:

$$\vartheta_i = \cos^{-1}\left(\frac{(C_{i+1}(x,y,z)-C_i(x,y,z))\cdot(C_{i+2}(x,y,z)-C_{i+1}(x,y,z))}{\|C_{i+1}(x,y,z)-C_i(x,y,z)\|\cdot\|C_{i+2}(x,y,z)-C_{i+1}(x,y,z)\|}\right), \vartheta_i \in \vartheta^j, j \in [1, n_L] \quad (26)$$

where $\vartheta^j$ is the turning angle set of the $j$-th slicing layer. As a result, sharp corner refers to turning point whose turning angle $\vartheta_i$ (rad) is larger than the critical angle $\delta_\theta$.

The essence of our VTG method is to reduce the sharp corners by point densification, and keep the vested trajectory unchanged. The second-order Taylor interpolation method is adopted as the basis of the proposed VTG for point densification, which is denoted as below:

$$u_{i+1} = u_i + \frac{L_i}{\|C'(u_i)\|} - \frac{\langle C'(u_i), C''(u_i)\rangle L_i^2}{2\|C'(u_i)\|^4} \qquad i = 1, \ldots, n \qquad (27)$$

where $L_i$ is interpolation distance, $\|\cdot\|$ is the operator to calculate Euclidean modulus,

$\langle \cdot, \cdot \rangle$ is the operator to calculate dot product of two vectors, and $n$ is the total number of interpolation points.

Besides, the speed fluctuation $\varepsilon$ is used to directly denote the vibration amplitude of toolpath, which is defined as follows:

$$\varepsilon = \frac{L_i - \|C(u_{i+1}) - C(u_i)\|}{L_i} \times 100\% \tag{28}$$

The definition of speed fluctuation signifies that the lower speed fluctuation represents the better kinematic performance. Hence, the extra kinematic limitations are exerted by the proposed VGT as below:

$$v_i \leq v_{max} \text{ AND } \begin{cases} a_{in} \leq A_n \\ a_{it} \leq A_t \end{cases} \text{ AND } \begin{cases} j_{in} \leq J_n \\ j_{it} \leq J_t \end{cases} \tag{29}$$

where $v_i$ (m·s$^{-1}$), $a_{in}$ (m·s$^{-2}$), $a_{it}$ (m·s$^{-2}$), $j_{in}$ (m·s$^{-3}$), $j_{it}$ (m·s$^{-3}$) are feed speed, normal acceleration, tangential acceleration, normal jerk, and tangential jerk, respectively. $v_{max}$ (m·s$^{-1}$), $A_n$ (m·s$^{-2}$), $A_t$ (m·s$^{-2}$), $J_n$ (m·s$^{-3}$), $J_t$ (m·s$^{-3}$) are maximal limitations of feed speed, normal acceleration, tangential acceleration, normal jerk, and tangential jerk allowed by the LDAM system, respectively.

Further, the maximal curvature $K_{cr}$ is defined to limit the curvature of toolpath which is generated by the proposed VGT. The maximal curvature $K_{cr}$ is expressed as follows:

$$K_{cr} = \min\left(\frac{8\delta}{(v_{max}T_s)^2 + 4\delta^2}, \frac{A_n}{v_{max}^2}, \sqrt{\frac{J_n}{v_{max}^3}}, \frac{1}{R_n}\right) \tag{30}$$

where $T_s$ is the time of interpolation cycle, and $R_n$ is the radius of printing nozzle (mm).

Algorithm 1 shows the pseudo-code of the proposed VTG. $\varphi$ and $\tau$ are set to control the convergence speed of the proposed VTG and the length of the interpolation segment, respectively. The distance between the new generated points $P_i$ and $P_{i+1}$ is exactly equal to the interpolation step $L_i$. Moreover, since $K_{cr}$ is limited by various kinematic parameters, the VTG enables the interpolated curvature to be lower than maximal curvature $K_{cr}$, and remains the kinematic parameters under the reasonable range allowed in LDAM system.

To evaluate the performance of the VGT, chord error $e_c$ is used as a crucial metric, which denotes the max distance between a chord and its corresponding arc length as

below:

$$e_c = r_i - \sqrt{r_i^2 - ((\|C_{i+1}(x,y,z) - C_i(x,y,z)\|)/2)^2} \tag{31}$$

where $r_i$ is the radius of the $i$-th circumscribed circle that constructed by three continuous points including $C_i(x,y,z), C_{i+1}(x,y,z), C_{i+2}(x,y,z)$.

---

**Algorithm 1**: Vibration-suppressed Toolpath Generation (VTG)

---

**Input:** $\varphi, \tau, u_0, L_i, K_{cr}$
**Output:** Interpolation point $P_{i+1}$

1: **while** $u_i \leq 1$ **do**
2: $\quad u_{i+1}^0 \leftarrow u_i + \frac{L_i}{\|C'(u_i)\|} - \frac{\langle C'(u_i), C''(u_i) \rangle L_i^2}{2\|C'(u_i)\|^4};$
3: $\quad$ **do**
4: $\quad\quad \Delta L_{i+1}^{(j)} \leftarrow \left\| C\left(u_{i+1}^{(j)}\right) - P_i \right\|;$
5: $\quad\quad F\left(u_{i+1}^{(j)}\right) \leftarrow \left\| C\left(u_{i+1}^{(j)}\right) - P_i \right\| - (1-\varphi)\Delta L_{i+1}^{(j)};$
6: $\quad\quad F'\left(u_{i+1}^{(j)}\right) \leftarrow \frac{\langle C(u_{i+1}^{(j)}) - P_i, C'(u_{i+1}^{(j)}) \rangle}{\|C(u_{i+1}^{(j)}) - P_i\|};$
7: $\quad\quad u_{i+1}^{(j+1)} \leftarrow u_{i+1}^{(j)} - \frac{F(u_{i+1}^{(j)})}{F'(u_{i+1}^{(j)})};$
8: $\quad\quad j \leftarrow 0,1,2,\cdots;$
9: $\quad$ **while** $\tau L_i < \Delta L_{i+1}^{(j)} < L_i$ and $K_{i,i+1} < K_{cr}$
10: $\quad u_{i+1} \leftarrow u_{i+1}^{(j)};$
11: $\quad P_{i+1} \leftarrow P_i + \frac{C(u_{i+1}) - P_i}{\|C(u_{i+1}) - P_i\|} L_i;$
12: **end while**
13: **return** $P_{i+1}$

---

## 4. Numerical experiment verification

As shown in Fig. 3, the W16 cylinder is adopted as an experimental model to verify the proposed VTG method, which owns 16 (4*4) independent cylinders. The ignition sequence of the experimental block model follows $C_1\ C_{14}\ C_9\ C_4\ C_7\ C_{12}\ C_{15}\ C_6\ C_{13}$ $C_8$ $C_3$ $C_{16}$ $C_{11}$ $C_2$ $C_5$ $C_{10}$, as shown in Fig. 3(a). In terms of the layer-wise fabrication manner of AM, the experimental block model is supposed to be sliced to multiple layers, enabling the LDAM machine to fabricate it layer-upon-layer. Fig. 3(b) shows the sliced region generated by a slicing plane, indicating the experimental block model demands the higher kinematic and energy performance to meet its sophisticated working characteristics. Examples of six different infill patterns for the experimental block model are shown in Fig. 4.

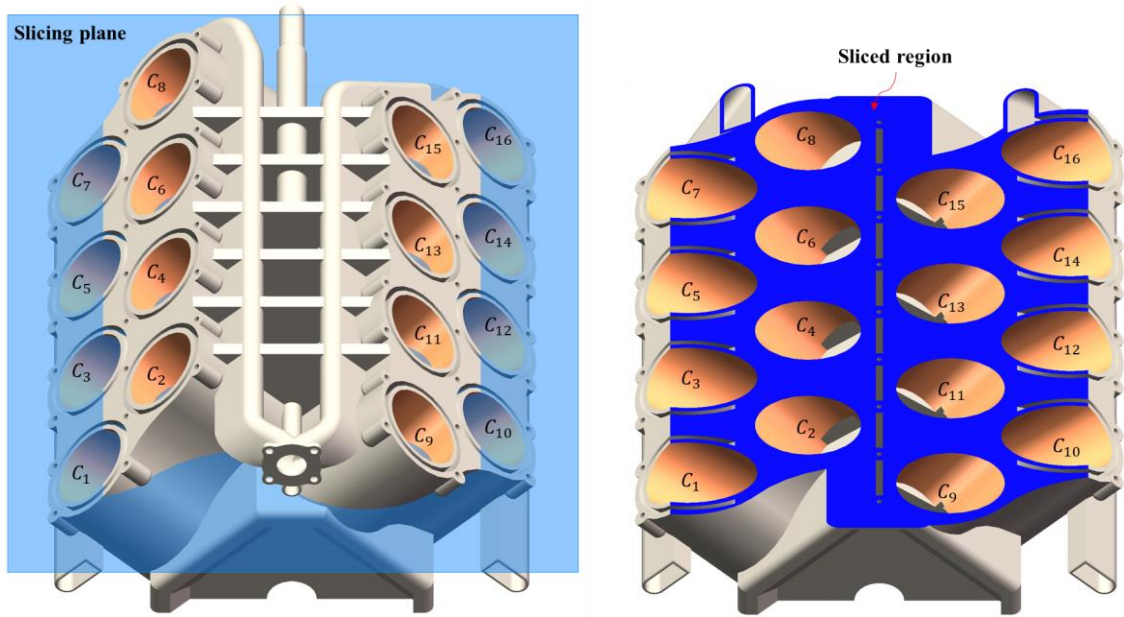

**Fig. 3** (a) W16 cylinder with a slicing plane; (b) Sliced region of the experimental block model.

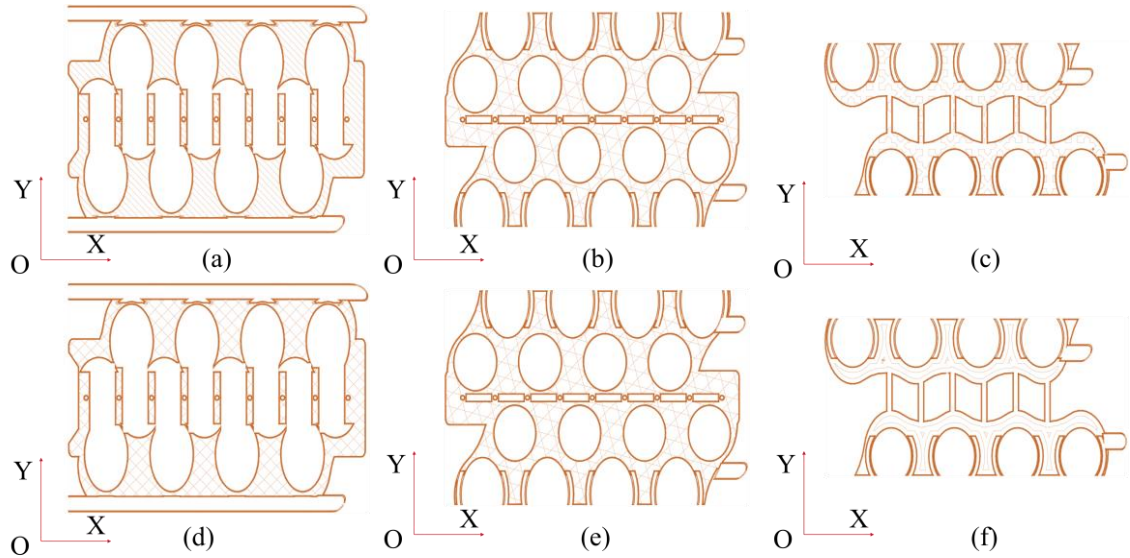

**Fig. 4** Examples of six different infill patterns for the experimental block model. (a) and (d) are line and grid respectively, which are located in $h_n =25\%$. (b) and (e) are triangle and tri-hexagon respectively, which are located in $h_n =50\%$. (c) and (f) are cross-line and concentric respectively, which are located in $h_n =75\%$.

Fig. 5 shows the comparsion results of layered trajectory length and mean angle for the experimental block model with diverse infill patterns. It can be seen from Fig. 5 (a) that the concentric pattern yields the minimal layered trajectory length than other infill patterns, and these patterns generate similar layered trajectory mean angle, as shown in Fig. 5 (b). The concentric pattern is thus adopted to conduct the following experiments for kinematic and energy performance optimization.

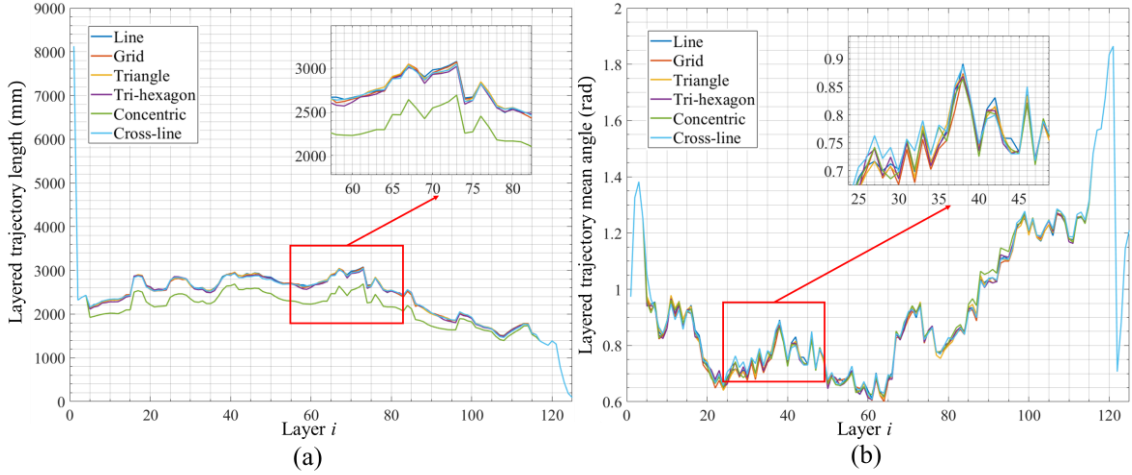

**Fig. 5** Comparsion results of layered trajectory length and mean angle for the experimental block model with diverse infill patterns.

Fig. 6 shows theoretical and actual toolpath points on the concentric pattern by using the proposed VTG method. In Fig. 6 (a), the maximum toolpath deviation is 0.1370 mm, located at (114.7211,83.3348), and the average value is 0.0865 mm. In Fig. 6 (b), the maximum toolpath deviation is 0.1381 mm, located at (118.0361,177.7574), and the average value is 0.0776mm. In Fig. 6 (c), the maximum toolpath deviation is 0.1387mm, located at (227.4984,45.6667), and the average value is 0.0699 mm. In Fig. 6 (d), the maximum toolpath deviation is 0.1398 mm, located at (143.6057,115.6550), and the average value is 0.0706 mm. In summary, Table 1 shows the comparisons for toolpath parameters before and after using the VTG method. The mean value of chord error is decreased as high as 22.04%, demonstrating that the proposed VGT method indeed helps the improvement of the kinematic performance for LDAM.

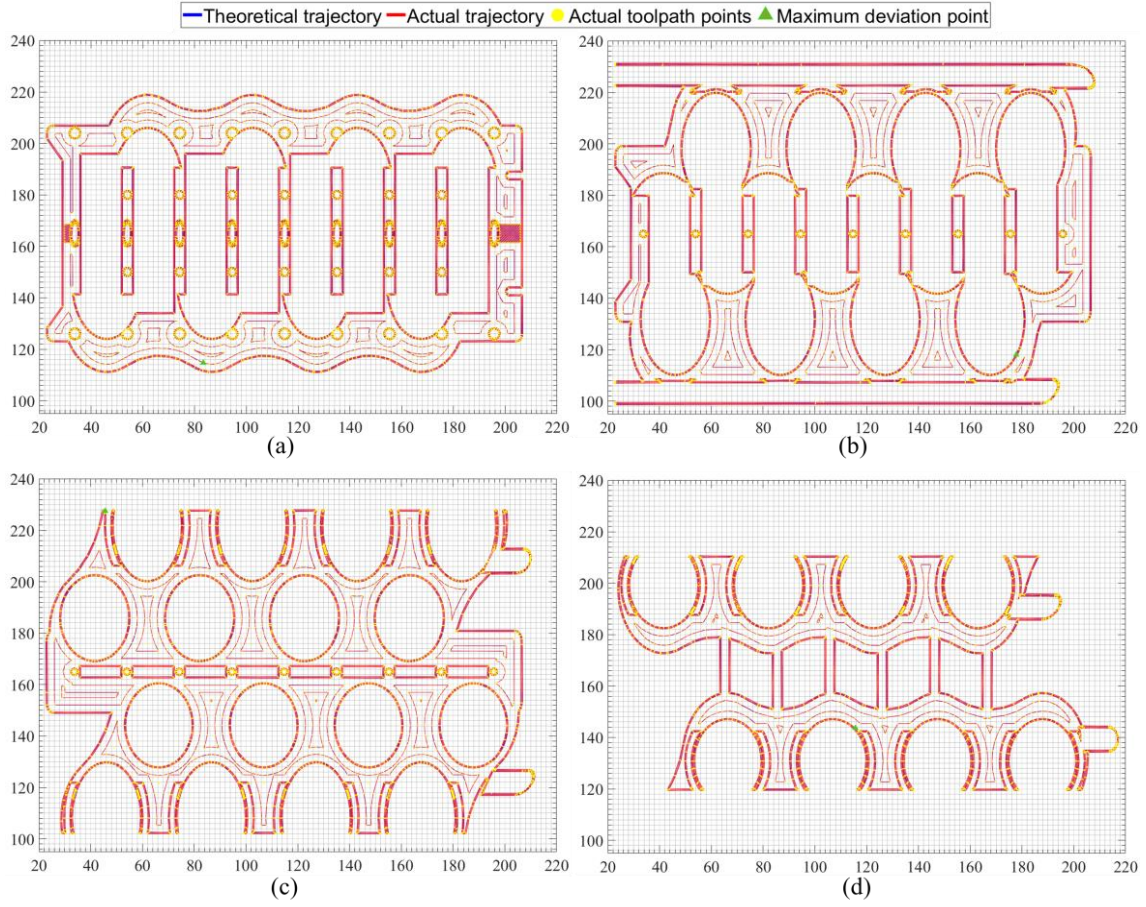

**Fig. 6** Theoretical and actual toolpath points on the concentric pattern by using the proposed VTG method, where (a) (b) (c) (d) are located in $h_n = 15\%, 25\%, 50\%, 75\%$, respectively.

**Table 1** Comparisons for toolpath parameters before and after using the VTG method.

| Metrics | Before | After | Ratio |
|---|---|---|---|
| Total toolpath length $L_{total}$ (mm) | 17256.96 | 14741.66 | -14.56% |
| Amount of toolpath points $n_{total}$ | 7647 | 6909 | -9.65% |
| Length of infill toolpath $L_{infill}$ (mm) | 4094.45 | 2829.49 | -30.89% |
| Amount of infill points | 880 | 871 | -0.99% |
| $n_{infill}$ | | | |
| Amount of turning points $n_{turning}$ | 360 | 211 | -41.50% |
| Mean value of chord error mean($e_c$) (mm) | 0.16 | 0.13 | -22.04% |

## 5. Physical experiment verification

### 5.1 Vibration measurement and analysis for LDAM

As shown in Fig. 7, we assemble a customized large dimension 3D printer and use it to conduct physical fabrication experiments to validate the proposed VTG method for LDAM. The dimension of the customized large dimension 3D printer is 1200mm x 1000mm x 650mm, and its printing dimension range could reach 1000mm x 800mm x 350mm. To real-time monitor and record vibration data, the multi-channel vibration recorder is used, whose vibration measuring range is $0.5 \pm 199.9$ m/$s^2$, measuring resolution is 0.1 m/$s^2$, working current is 12 ~ 35 mA, frequency range is 10 Hz~1KHz, communication mode is RS232/TTL, working humidity is less than 85%, RH operating voltage is DC 9 ~ 12 V, and working temperature is 0 °C ~ + 50 °C. Dual-channel vibration sensors are respectively fixed on the nozzle and working table, to measure and record the vibration signals of the nozzle and working table, *i.e.*, $Vib_n$ and $Vib_e$. The commercial PLA is used as the raw material to fabricate parts in the large dimension 3D printer. Fig. 8 shows graphical user interface (GUI) of real-time XYZ axis acceleration monitoring software. The electrical power analyzer, HPM-series (as shown in Fig. 7-16), is utilized to measure and record energy consumption during the physical experiments, to evaluate the energy performance in LDAM.

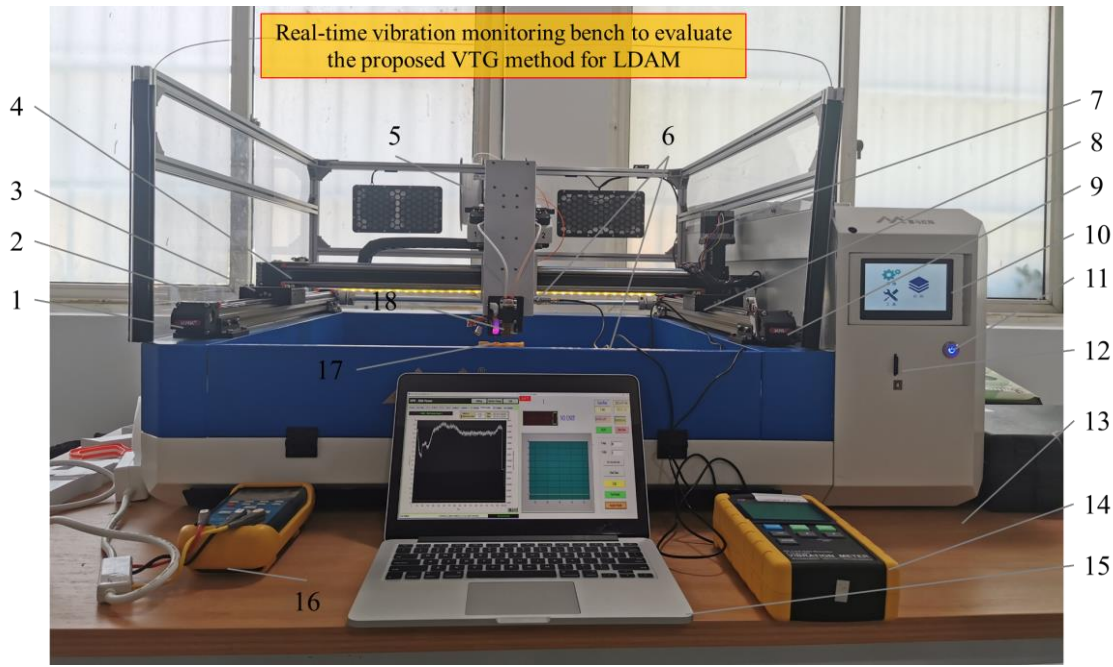

**Fig. 7** Customized large dimension 3D printer to conduct physical fabrication experiments. 1-DC servo motor in X-axis; 2-X-axis slider (left); 3-Slave pedestal (left); 4-Y-axis crossbeam; 5-Filament; 6-Dual channel vibration sensor; 7-DC servo motor in Z-axis; 8-Slave pedestal (right); 9-X-axis slider (right); 10-Control panel; 11-Power button; 12-SD card slot; 13-Experimental bench; 14-Multi-channel vibration measuring instrument; 15-Computer for data display; 16-Electrical power analyzer for energy efficiency monitoring; 17-Printing workpiece; 18-Nozzle.

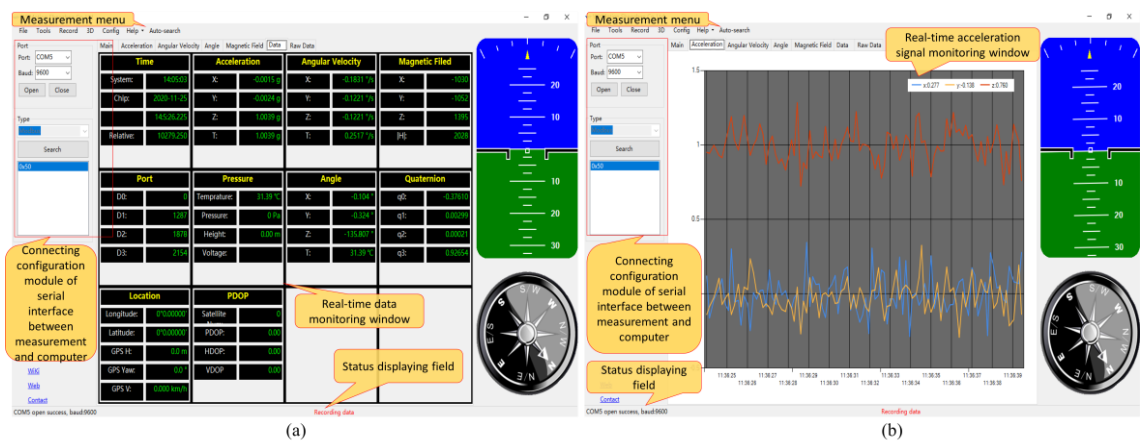

**Fig. 8** Graphical user interface (GUI) of real-time XYZ axis acceleration monitoring software.

Fig. 9 shows the measured XYZ axis accelerations $\ddot{x}_4$ which are recorded during the manufacturing process before and after optimization, to verify the efficiency of the proposed VTG method. Compared with Fig. 9 (a) and Fig. 9 (b), it can be seen that the XYZ axis acceleration amplitude after optimization is obviously decreased as expected. Fig. 9 (c) shows the layered vibration amplitude in XYZ axis before and after optimization, and the optimized vibration amplitude is lower than unoptimized vibration amplitude, demonstrating that mechanical vibration can be decreased by the proposed VTG method. Furthermore, Fig. 9 (d) shows the relative rate of layered vibration amplitude after optimization, where the relative rates in most of layers are negative, indicating that the proposed VGT results in a steady decrease of mechanical vibration. These results prove that the proposed VGT is able to suppress mechanical vibration by toolpath generation for kinematic performance optimization in LDAM.

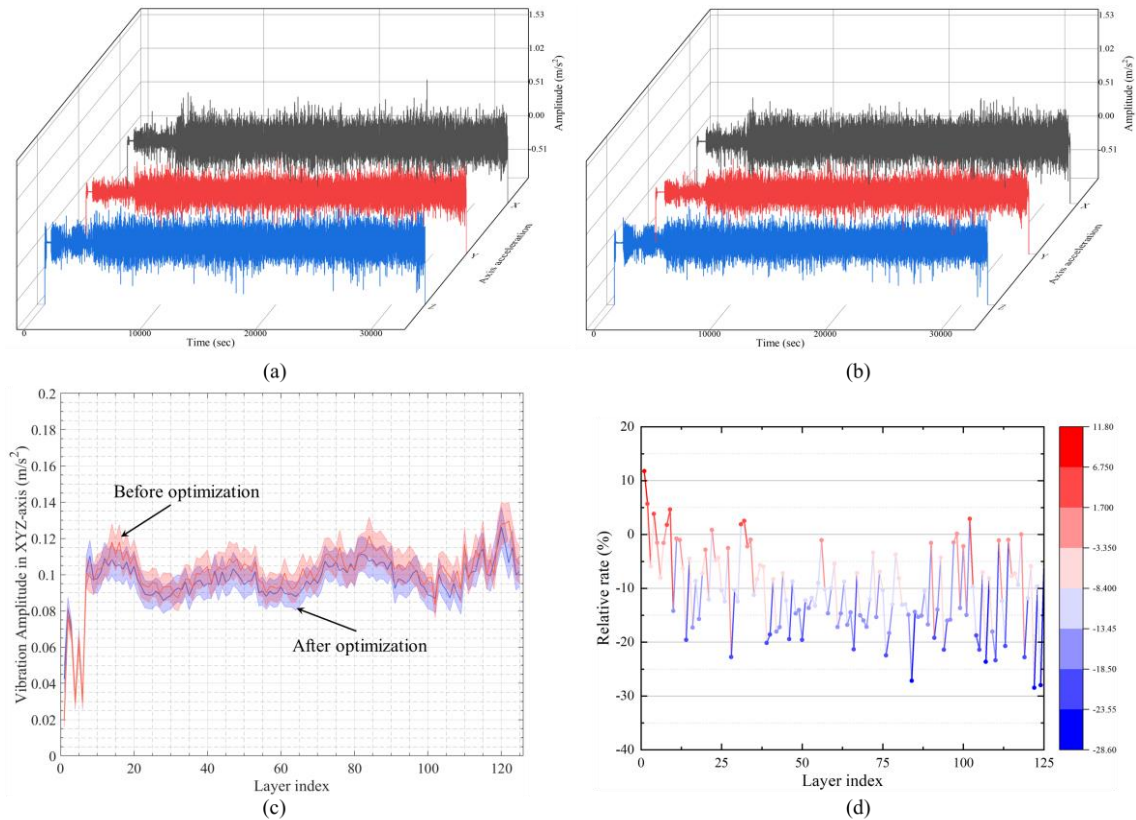

**Fig. 9** (a) Real-time XYZ axis acceleration amplitude before optimization; (b) Real-time XYZ axis acceleration amplitude after optimization; (c) Layered vibration amplitude in XYZ axis before and after optimization; (d) Relative rate of layered vibration amplitude after optimization.

For further power spectrum density (PSD) analysis, the sampling frequency is set to 200 Hz, and the Hamming window size is set to 512. Fig. 10 shows PSD analysis of mechanical vibration in working table before and after optimization, with the vibration amplitude in time domain. From Fig. 10 (a) and Fig. 10 (b), it can be seen that the PSD of mechanical vibration in working table is mostly located in the low frequency region, where the range of 5 ~ 40Hz accounts for most of high amplitude vibration. Also, the values of the top-3 maximum power spectrum is obviously decreased by using the proposed VTG, demonstrating that our method achieves the better kinematic performance improvement gains in both time and spectrum domains.

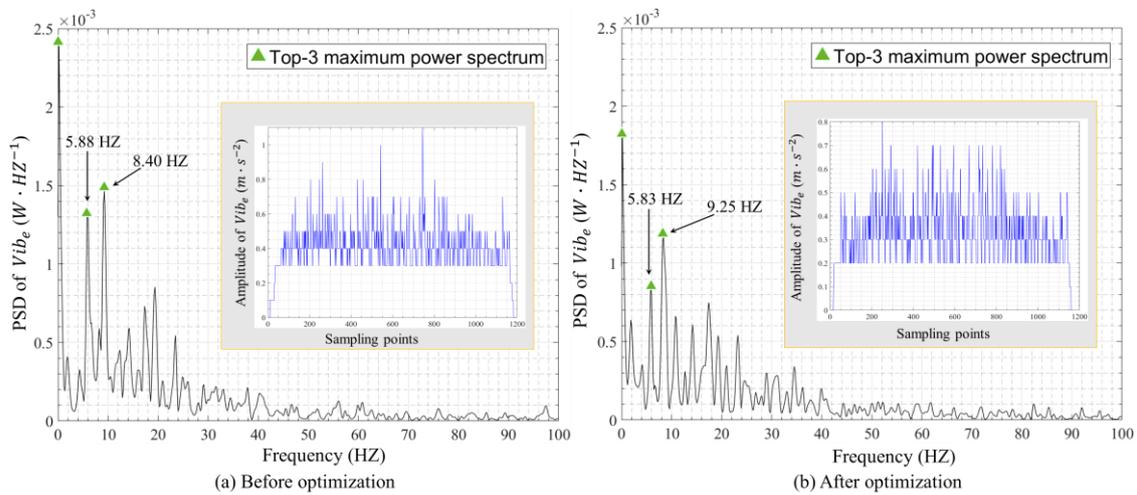

**Fig. 10** (a) PSD analysis of mechanical vibration in working table before optimization; (b) PSD analysis of mechanical vibration in working table after optimization.

### 5.2 Surface roughness measurement

Surface roughness is a common metric to evaluate the smoothness of the manufactured surface. The smaller surface roughness represents the smoother surface. Hence, we use a non-contact optical profilometer (veeco/NT9100) to measure the surface roughness of the fabricated specimen, where its range of the objectives magnification is 0.75X~100X, measurement array is 640x480, range of the vertical measurement is 0.1nm~10nm, and optical resolution is 0.49μm. Fig. 11 shows the non-contact optical profilometer for surface roughness measurement of the fabricated specimen.

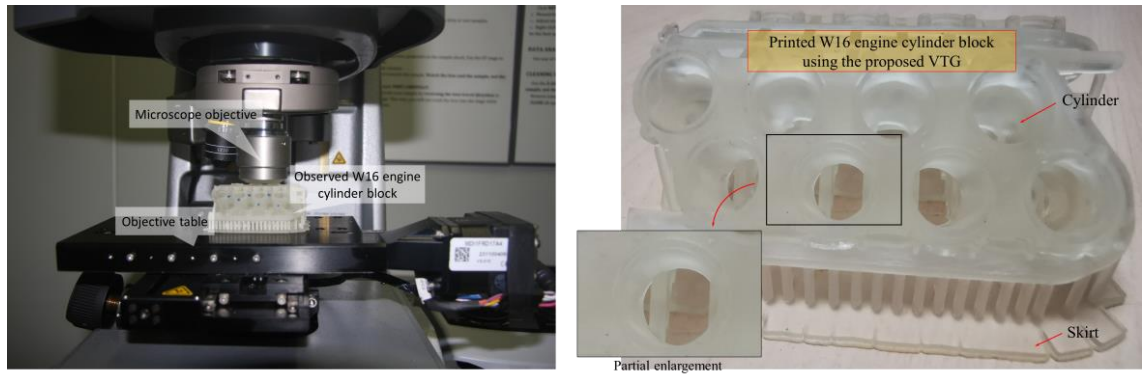

**Fig. 11** The non-contact optical profilometer for surface roughness measurement of the fabricated specimen.

Fig. 12 shows the surface topography images which are measured by the non-contact optical profilometer where (a) (b) are measured in two different regions without using the proposed VGT method, and (c) (d) are measured in two corresponding regions by using the proposed VGT method. Compared with Fig. 12 (a) and Fig. 12 (b), Fig. 12 (c) and Fig. 12 (d) shows their surface roughness (Ra) are decreased to 326.50 nm and 407.58 nm, respectively, signifying that the surface quality also benefits from the proposed VTG method with an improved kinematic performance simultaneously.

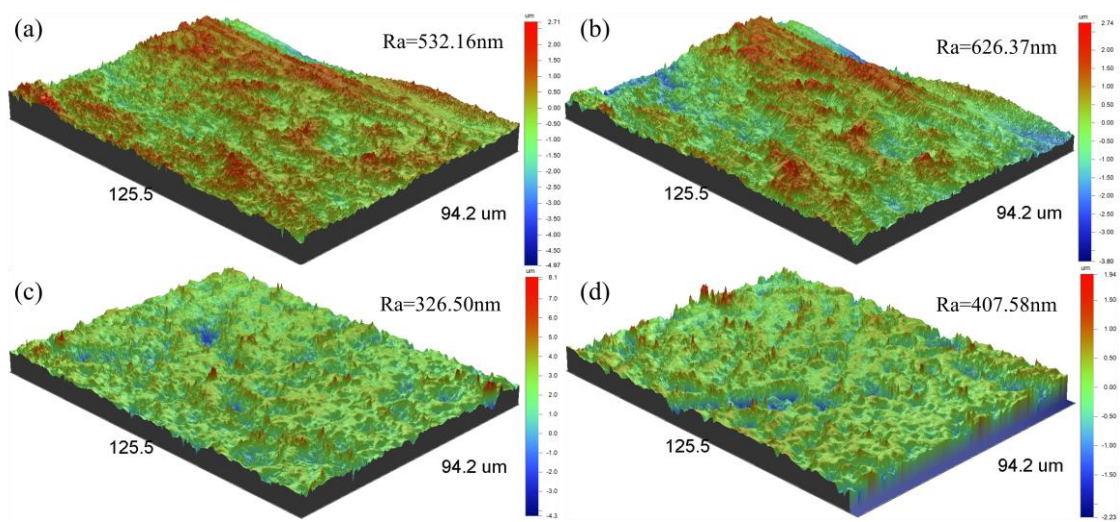

**Fig. 12** Surface topography images which are measured by the non-contact optical profilometer,

where (a) (b) are measured without using the proposed VGT method, and (c) (d) are measured by using the proposed VGT method.

## 5.3 Comparisons of the kinematic and energy performance

To summary the improvement gains by using our approach, Table 2 compares various metrics for evaluating kinematic and energy performance improvement before and after using the proposed VTG. The result illustrates that the mean toolpath deviation and total energy consumption $E$ are decreased by -11.98% and 12.05%, respectively, demonstrating that our VGT is able to simultaneously achieve kinematic and energy performance improvement in LDAM, even though the fabricated part owns aberrant and complex morphology.

**Table 2** Various metrics for evaluating kinematic and energy performance improvement before and after using the proposed VTG method.

| Optimization aims | Metrics | Before | After | Ratio |
|---|---|---|---|---|
| Kinematic performance | Maximal toolpath deviation (mm) | 0.1676 | 0.1353 | -19.27% |
| | Minimal toolpath deviation (mm) | 0.0018 | 0.0015 | -12.71% |
| | Mean toolpath deviation (mm) | 0.0874 | 0.0769 | -11.98% |
| Energy performance | Total energy consumption $E$ (J) | 9.378e+05 | 8.248e+05 | -12.05% |
| | Material consumption $m_{mater}$ (g) | 41.9 | 34.0 | -18.90% |
| | Carbon emission $m_{CO_2}$ (g) | 196.0 | 178.6 | -8.87% |

## 6. Conclusions

A Kinematic modeling of LDAM system is constructed by using the equivalent lumped mass model, revealing the relationship between the input force and output displacements in LDAM system. The energy consumption modeling related to LDAM kinematics is thus put forward, signifying that the energy of servo system accounts for main part of the total energy consumption in LDAM. In light of the above-established kinematic model and the corresponding energy consumption model, the vibration-suppressed toolpath generation (VTG) method is proposed to customize servo trajectory for kinematic and energy performance optimization in LDAM. The VTG enables the interpolated curvature to be lower than pre-defined maximal curvature, and remains the kinematic parameters under the reasonable range allowed in LDAM system.

Extensive numerical and physical experiments are conducted on the W16 cylinder to evaluate the effectiveness of our VGT for LDAM. To be concrete, we measure and analyze the mechanical vibration amplitude, and assess the surface roughness before and after using the proposed VGT. These results illustrate that the mean toolpath deviation and total energy consumption are decreased by -11.98% and 12.05%, respectively, demonstrating that our VGT is able to simultaneously achieve kinematic and energy performance optimization in LDAM systems, even though the fabricated part owns aberrant and complex morphology.